\documentclass{jetpl}
\twocolumn

\usepackage{amsmath,amssymb,epsfig,color,pstricks,bm,alltt}

\usepackage[colorlinks,plainpages=false,linkcolor=black,urlcolor=blue,citecolor=black,pdfpagemode=UseNone,pdfstartview=FitBH]{hyperref}

\definecolor{MyDarkBlue}{rgb}{0,  0.3,  0.9}
\definecolor{MyDarkBlack}{rgb}{0,  0,  0}

\begin{document}

\lat

\title{Electronic structure of NaFeAs superconductor: LDA+DMFT 
calculations compared with ARPES experiment}

\rtitle{Electronic structure of NaFeAs superconductor: LDA+DMFT vs. ARPES}

\sodtitle{Electronic structure of NaFeAs superconductor: LDA+DMFT calculations 
compared with ARPES experiment}

\author{$^a$I.\ A.\ Nekrasov\thanks{E-mail: nekrasov@iep.uran.ru}, $^a$N.\ S.\ Pavlov\thanks{E-mail: pavlov@iep.uran.ru},
$^{a,b}$M.\ V.\ Sadovskii\thanks{E-mail: sadovski@iep.uran.ru}}

\rauthor{I.\ A.\ Nekrasov, N.\ S.\ Pavlov, M.\ V.\ Sadovskii}

\sodauthor{I.\ A.\ Nekrasov, N.\ S.\ Pavlov, M.\ V.\ Sadovskii }

\sodauthor{I.\ A.\ Nekrasov, N.\ S.\ Pavlov, M.\ V.\ Sadovskii}

\address{$^a$Institute for Electrophysics, Russian Academy of Sciences, 
Ural Branch, Amundsen str. 106,  Ekaterinburg, 620016, Russia\\
$^b$ M.N. Mikheev Institute for Metal Physics, Russian Academy of Sciences, Ural Branch,
S.Kovalevskoi str. 18, Ekaterinburg, 620290, Russia
}


\abstract{We present the results of extended theoretical LDA+DMFT calculations for 
a new iron-pnictide high temperature superconductor NaFeAs compared with the 
recent high quality angle-resolved photoemission (ARPES) experiments on this system [1]. 
The universal manifestation of correlation effects in iron-pnictides is narrowing 
of conducting bands near the Fermi level. Our calculations demonstrate that for NaFeAs 
the effective mass is renormalized on average by a factor of the order of 3, 
in good agreement with ARPES data. This is essentially due to correlation effects 
on Fe-3d orbitals only and no additional interactions with with any kind of Boson modes, 
as suggested in [1], are necessary to describe the experiment.
Also we show that ARPES data taken at about 160 eV beam energy most probably corresponds to 
$k_z=\pi$ Brillouin zone boundary, while ARPES data measured at about 80 eV beam energy 
rather represents $k_z=0$. Contributions of different Fe-3d orbitals into spectral function map 
are also discussed.}


\maketitle

\section{Introduction}

The family of iron based high-temperature superconductors 
first discovered in 2008 \cite{kamihara_08} 
still attracts a lot of scientific attention. Experimental and theoretical works 
on these materials are now discussed in several extended reviews~\cite{UFN,Hoso_09,Johnson,Mazin}.
Detailed comparison of electronic band structures of iron pnictides and iron halcogenides,
together with some related compounds was given in Refs. \cite{PvsC,relcomp}. 

One of the classes of iron pnictides is the so called 111 system with
parent compound Li$_{1-x}$FeAs with T$_c$=18 K~\cite{cryst,wang_4688}.
LDA band structure of the LiFeAs was first described in the Refs. \cite{Nekr3,Shein2}.

One of the most effective experimental techniques to probe electronic band
structure of these and similar systems is the angle-resolved photoemession spectroscopy 
(ARPES) \cite{Damascelli}. A review of the present day status of ARPES results for iron 
based superconductors can be found in Ref. \cite{Kordyuk}.

Soon after the discovery of iron based superconductors it was shown both experimentally
\cite{Popovich, Sergei_LiFeAs, Ding, Cui_NFCA, Orbitalgap} (mainly by ARPES)
and theoretically \cite{Haule,Craco,Shorikov,Ba122_DMFT} (within the LDA+DMFT hybrid 
computational scheme \cite{LDADMFT}) that electronic correlations on Fe sites are essential 
to describe the physics in these materials. The main manifestation of correlations is 
simple narrowing (compression) of LDA bandwidth near the Fermi level by the factor of the 
order from 2 to 4. At the same time the topology of ARPES determined Fermi surfaces is quite 
similar to those obtained from simple LDA calculations, showing two or three hole cylinders 
around $\Gamma$--point in the Brillouin zone and two electron Fermi surface sheets around 
($\pi,\pi$) point.

This work was inspired by recent high quality ARPES data for NaFeAs system \cite{Evtushinsky_14}
and is devoted to the detailed comparison of these results with LDA+DMFT calculations of
electronic structure of this system, showing rather satisfactory agreement with these
experiments. Thus, only the account of electronic correlations is sufficient to explain the
major features of electronic spetrum of NaFeAs, and there is no need for any additional
interactions with any kind of Boson modes (as was suggested in Ref. \cite{Evtushinsky_14}).

\section{Electronic structure}

\begin{figure}
\includegraphics[clip=true,width=0.45\textwidth]{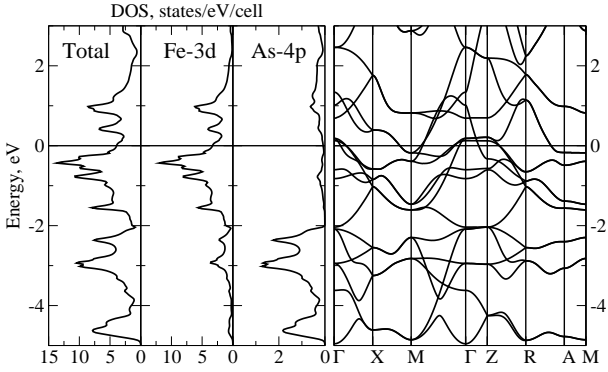}
\caption{Fig. 1. LDA calculated band dispersions (right) and densities of 
states (left) of paramagnetic NaFeAs. The Fermi level $E_F$ is at zero energy.} 
\end{figure}

The crystal structure of NaFeAs has tetragonal structure with the space group $P$4/$nmm$
and lattice parameters $a=3.9494$~\AA, $c=7.0396$~\AA.
The experimentally obtained crystallographic positions are the following Fe(2b)
(0.75, 0.25, 0.0), Na(2c)  (0.25, 0.25, z$_{Na}$),
As(2c) (0.25, 0.25, z$_{As}$), z$_{As}$=0.20278,
z$_{Na}$=0.64602~\cite{NaFeAs_cryst}. That is quite similar to LiFeAs crystal structure \cite{cryst,Nekr3}.

In Fig.~1 we show LDA band dispersions (on the right) and densities of states (DOS) (on the left)
calculated within FP-LAPW method \cite{wien2k}. Bands in the vicinity of the Fermi level
have predominantly Fe-3d character and are essentially similar to the previously studied case of LiFeAs
described elsewhere \cite{Nekr3,Shein2}. The As-4p states belong to the -2 to -5 eV energy interval.

To perform DMFT part of LDA+DMFT calculations we used CT-QMC  impurity solver \cite{ctqmc,triqs}.
In order to link LDA and DMFT we exploited Fe-3d and As-4p projected Wannier functions LDA Hamiltonian 
for about 1500 $k$-points. Standard  wien2wannier interface \cite{wien2wannier} and wannier90 projecting technique 
\cite{wannier90} were applied to this end. The DMFT(CT-QMC) computations were done at reciprocal temperature $\beta=40$ 
with about 10$^7$ Monte-Carlo sweeps. Hubbard model interaction parameters were taken to be $U$=3.5 eV and $J$=0.85 eV 
as typical values for pnictides in general and close NaFeAs relative -- LiFeAs in particular 
\cite{Skornyakov_lifeas,Jeschke_12,Kotliar_12}.

\begin{figure}
\includegraphics[clip=true,width=0.45\textwidth]{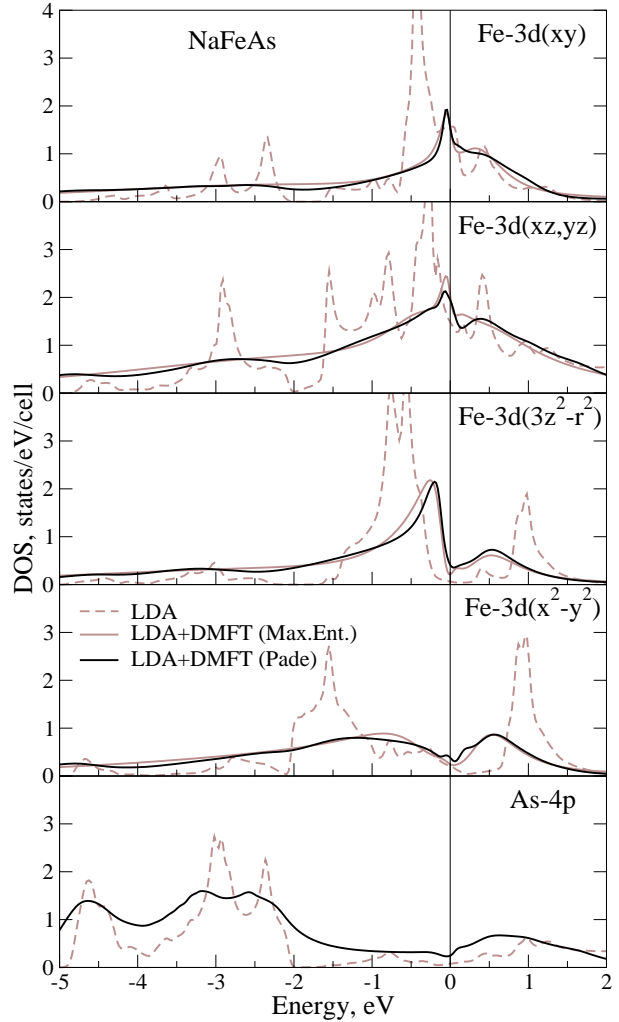}
\caption{Fig. 2. Comparison of orbital resolved densities of states for Fe-3d shell of NaFeAs
obtained within LDA (gray dashed line) and LDA+DMFT (solid gray and black lines). Zero energy is the Fermi level.} 
\end{figure}

Fig. 2 shows the comparison of orbital resolved densities of states for Fe-3d shell of NaFeAs
obtained within LDA (gray dashed line) and LDA+DMFT (solid gray and black lines).
Solid gray and black lines show LDA+DMFT densities of states obtained by different methods of analytic continuation.
Gray lines are obtained directly from DMFT(CT-QMC) Green function G($\tau$) by maximum entropy method \cite{ME}.
Overall lineshapes of LDA+DMFT densities of states are identical to those already published in the literature
for LiFeAs \cite{Skornyakov_lifeas,Jeschke_12,Kotliar_12} and NaFeAs \cite{Kotliar_12}.
Most affected by correlations are Fe-3d(t$_{2g}$) orbitals $xy$ and degenerate $xz,yz$. These orbitals
form narrow pronounced peaks near the Fermi level. On the other hand Fe-3d(e$_{g}$) orbitals $3z^2-r^2$ 
and $x^2-y^2$ just remind the broadened LDA densities of states.

\begin{figure}
{\center
\includegraphics[clip=true,width=0.35\textwidth]{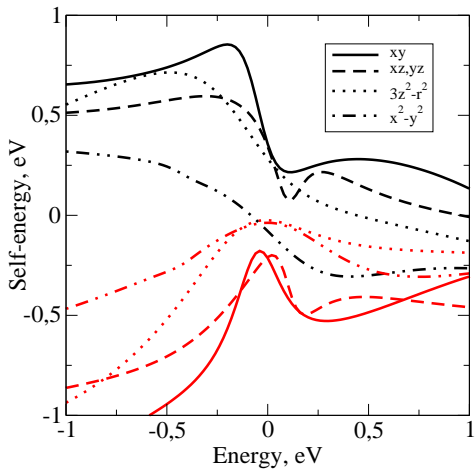}
}
\caption{Fig. 3. LDA+DMFT calculated self-energies for different Fe-3d orbils of NaFeAs near the Fermi level.
Black lines -- real part, gray lines -- imaginary part.
The Fermi level $E_F$ is at zero energy. } 
\end{figure}

To produce LDA+DMFT spectral function maps for direct comparison with ARPES data we need to know the local
self-energy $\Sigma(\omega)$. To find it we have to perform analytic continuation from Matsubara frequencies 
to real ones. To this end we have applied Pade approximant algorithm \cite{pade}.
The fact that both gray and solid lines coinside well in Fig. 2 tells us, that this analytic continuation 
is done rather satisfactory. Corresponding self-energies for different Fe-3d orbitals near the Fermi level 
are shown on Fig. 3. From the real part of self-energy we can obtain the mass renormalization factor for different orbitals:
$m^*/m_{xy}\approx$3.8, $m^*/m_{xz,yz}\approx$3.9,  $m^*/m_{3z^2-r^2}\approx$2 and $m^*/m_{x^2-y^2}\approx$1.
These numbers agree well with variety of previous theoretical works for LiFeAs and NaFeAs  
\cite{Skornyakov_lifeas,Jeschke_12,Kotliar_12}. Thus only the account of local Coulomb correlations on the Fe sites is enough 
to produce such renormalization and no extra interaction with possible Boson mode is necessary in contrast to the
proposal of Ref. \cite{Evtushinsky_14}.

\begin{figure}
\includegraphics[clip=true,width=0.45\textwidth]{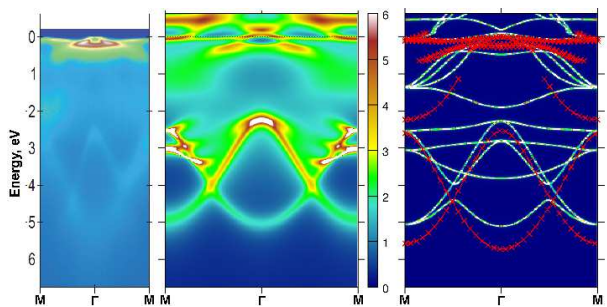}
\caption{Fig. 4. Comparison of experimental ARPES (left panel) [1] and LDA+DMFT (middle panel)
spectral functions in the M$\Gamma$M high symmetry direction for NaFeAs for the wide range of binding
energies containing Fe-3d and As-4p states. On the right panel maxima of experimental (crosses) [1] and 
theoretical (white lines) extracted from corresponding spectral functions are presented.
The Fermi level $E_F$ is at zero energy. } 
\end{figure}

Typically experimental ARPES data are presented in a rather narrow energy interval of few tenth of eV close to the 
Fermi level (for LiFeAs and NaFeAs see Refs. \cite{Kotliar_12,Borisenko_10,Yi_12,Cui_12,Fink_15}). 
However, in Ref. \cite{Evtushinsky_14} ARPES data were measured down to a quite large binding energies about 6 eV 
with rather high resolution allowing to extract different bands.

In Fig. 4 we compare experimental ARPES spectral functions for NaFeAs (left panel) \cite{Evtushinsky_14} 
along the the M$\Gamma$M high symmetry direction with LDA+DMFT calculated (middle panel) spectral function map for a wide energy window.
On both of these panels one can see rather high intensity region from 0 to 0.5 eV formed by quasiparticle bands
near the Fermi level and then from -2 to -5 eV we can observe As-4p bands. To compare experimental and theoretical bands 
dispersions on the right panel of Fig. 4 we plot the dispersions for the maxima of experimental (crosses) and theoretical (white lines)
spectral functions. 

According to Ref. \cite{Evtushinsky_14} ARPES bands line shapes remind very much the LDA bands,  compressed by an almost constant 
factor of the order of 3 for all energies. By analyzing the real part of self-energies from Fig. 3 we can convince ourselves,
that this correlation narrowing is essentially frequency dependent. Extended discussion of similar situation was given in 
our recent work on  KaFe$_2$Se$_2$ \cite{kfese_dmft}. Actually, the LDA bands located in the interval from -0.5 eV to 0.25 eV 
become more narrowed due to correlations. At larger energies,  the bands stay at about the same positions as in LDA or get 
more spread in energy since the slope of the real part of self-energy is changed to the positive one.

As to As-4p bands ARPES experiment resolves only 2 bands instead of 6 (2 As atoms in the unit cell).
Despite the general shape of the bands being quite similar in both cases, the experiment shows As-4p states about 
0.5 eV lower in energy than obtained in LDA+DMFT. This can be explained in the framework of generalized LDA'+DMFT 
calculations \cite{CLDA,CLDA_long}, which allows one a better description of Fe(3d)-As(4p) energy splitting, 
as was shown for example for KaFe$_2$Se$_2$ system \cite{kfese_dmft}. Indeed our LDA'+DMFT calculations showed, 
that As-4p states appeared about 0.5 eV lower in energy.

Between quasiparticle bands and As-4p bands there is a rather low intensity region (-0.5 eV -- -2 eV) seen in Fig. 4 
on the left and middle panels. First of all, it appears because there are almost no bands in this energy interval,
and secondly in this region we have a crossover from the well defined quasiparticle bands with quite low damping 
to the rest of the bands placed at higher binding energies.
This fact is illustrated by gray lines on Fig. 3, representing the imaginary parts $\Sigma''(\omega)$ of LDA+DMFT 
calculated self-energies for all Fe-3d orbitals. Near energy zero (Fermi level) $\Sigma''(\omega)$ is about 0.2  eV or 
less for all correlated states. At the same time real parts of the self-energies $\Sigma'(\omega)$ has negative slope 
near the Fermi level, which corresponds to well defined quasiparticles. Following $\Sigma'(\omega)$ behavior one can 
find that it has peak at about 0.25 eV, which corresponds to the end of quasiperticle region and
$\Sigma''(\omega)$ grows quite rapidly beyond this energy.
Nearly the same behavior of $\Sigma'(\omega)$ and $\Sigma''(\omega)$ was assumed in the Ref. \cite{Evtushinsky_14} 
and related to interaction with some ``unknown Boson mode'', distinguishing NaFeAs as unconventional superconductor. 
Again we claim that just the local Coulomb correlations on the Fe sites can do all that alone.

\begin{figure*}[!ht]
\includegraphics[clip=true,width=\textwidth]{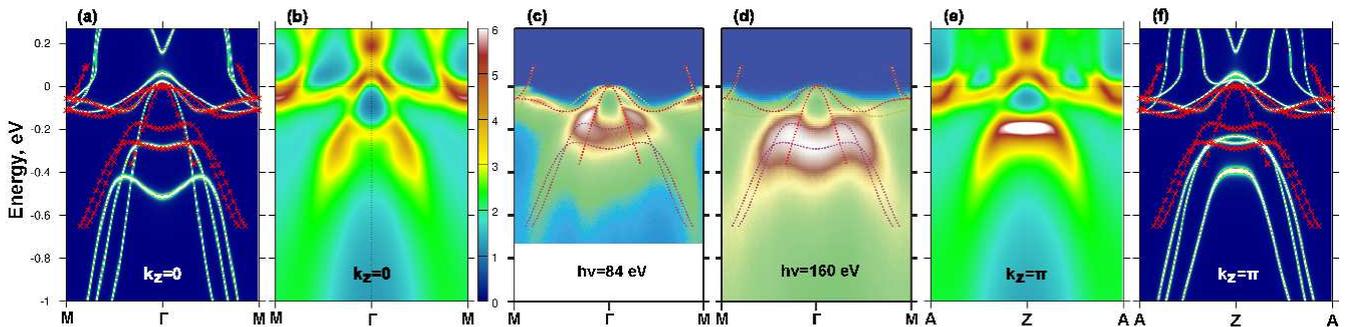}
\caption{Fig. 5. Comparison of experimental ARPES (panels c and d) [1] and LDA+DMFT (panels b and e)
spectral functions in the M$\Gamma$M and AZA high symmetry directions for NaFeAs near the Fermi level.
On the panels a and f experimental (crosses) [1] and theoretical (white lines) maxima dispersions 
of spectral functions are presented. The Fermi level $E_F$ is at zero energy. } 
\end{figure*}

Now we turn to the quasiparticle bands dispersions in the close proximity if the Fermi level.
Corresponding comparison of experimental ARPES data and theoretical LDA+DMFT spectral functions for NaFeAs is given
in Fig. 5. Here we show only the data for M$\Gamma$M high symmetry direction, the results for other symmetry drections
can be found in the Supplementary Material \cite{Suppl}. Experimental data were obtained at two different 
rather distinctive beam energies around 80 eV and 160 eV (see panels c and d on Fig. 5).
At bird eye view for both beam energies experimental picture looks similar, but in fact there are some remarkable 
differences. For 84 eV data $xy$ and $xz,yz$ bands close to the Fermi level are more intensive as compared to 160 eV data.
On the other hand $3z^2-r^2$ band at about -0.2 eV looks more intensive in 160 eV data.

To clarify this fact we suggest following explanation. It is well known that by varying the beam energy in ARPES experiments
one can access different values of $k_z$ component of the momenta \cite{Damascelli}. However to get the precise value of 
$k_z$ one should know the exact geometry of ARPES experiment \cite{Damascelli}, work function and inner potential for this 
particular material \cite{Brouet}. Since we do not know all these precisely,  we can try some speculations.
In Fig. 5 we plotted LDA+DMFT calculated spectral functions for  $k_z=0$ (panel b) and $k_z=\pi$ (panel e).
Now moving from $k_z=0$ (panel b) to $k_z=\pi$ (panel e) we can observe the same trend as one goes from 84 eV (panel c) 
to 160 eV (panel d) beam energy in the experiment.

Although iron based superconductors have pronounced layered structure still these systems are $quasi$ two-dimensional
and thereby possess some finite dispersion along $k_z$ axis. This fact is reflected on panels a and f of Fig. 5 where 
LDA+DMFT spectral function maxima dispersions (white lines) are shown at $k_z=0$ (panel a) and $k_z=\pi$ (panel f).
For the case of $k_z=\pi$ (panel f) $xz,yz$ bands are no more degenerate in $\Gamma$-point. One of $xz,yz$ bands branches 
goes down in energy to -0.2 eV and becomes degenerate with one of $3z^2-r^2$ bands. At the same time $3z^2-r^2$ band goes 
down to -0.4 eV at $\Gamma$-point and becomes more flat. All that, in contrast to the $k_z=0$ case, results in higher 
intensity of LDA+DMFT spectral function (panel e) around -0.2 eV and lower intensity at the Fermi level. 
The later one agrees better with 160 eV ARPES data, than with 84 eV ARPES data. Here one should stress that $xy$ band and 
one of  $xz,yz$ bands branches right below the Fermi level keep their shapes almost unchanged for both $k_z=0$ and $k_z=\pi$. 

Somewhat larger intensity of LDA+DMFT spectral functions in comparison with experiment near M point
arises because of quite strong  $xy$ contribution in this region. However in the ARPES data \cite{Evtushinsky_14}
$xy$ band is almost hidden, perhaps due to matrix elements effects (see Supplementary Material \cite{Suppl}).
Note also that shown experimental ARPES maxima (crosses on panels a and f) in accordance with
Ref. \cite{Evtushinsky_14} do not depend on beam energy.

\section {Conclusion}

In this paper we have presented the results of extended LDA+DMFT(CT-QMC) theoretical analysis of 
recent high quality angle-resolved photoemission (ARPES) experiments on a new iron-pnictide high 
temperature superconductor NaFeAs \cite{Evtushinsky_14}. The well known and rather universal manifestation 
of correlation effects in iron--pnictides is the renormalization (narrowing) of conducting bands near 
the Fermi level by a factor of 2 to 4. Corresponding mass renormalization factors for different orbitals
were obtained from LDA+DMFT calculations and, in our opinion, no extra interaction with some ``unknown Boson mode'' 
distinguishing NaFeAs as unconventional superconductor is necessary in contrast to the suggestion 
of Ref. \cite{Evtushinsky_14}. 

Also we have shown that ARPES data taken at 160 eV beam energy most probably corresponds to 
$k_z=\pi$ Brillouin zone boundary, while the data measured at about 80 eV beam energy reproduces
$k_z=0$. Theoretical analysis of spectral weight redistribution support this point of view. 
Comparison of different Fe-3d orbitals contributions to spectral function maps for
vertically and horizontally polarized ARPES data also favors the last statement (see Supplementary Material \cite{Suppl}).

We thank D. Evtushinsky for many helpful discussions of ARPES experimental data,
A. Lichtenstein and I. Krivenko for providing us their CT-QMC code and A. Sandvik
for making available his maximum entropy program. We are especially grateful to G.N. Rykovanov 
for providing us the access to VNIITF ``Zubr'' supercomputer, at which most of our CT-QMC 
computations were performed. Also part of CT-QMC computations were performed
at supercomputer ``Uran'' at the Institute of Mathematics and Mechanics UB RAS.

This work was done under the State Contract No. 0389-2014-0001 and partly supported 
by RFBR grant No. 14-02-00065.


\newpage

\begin{center}

{\bf Supplemental Material to ``Electronic structure of NaFeAs superconductor: 
LDA+DMFT calculations compared with ARPES experiment''}

\end{center}

In this Supplement we provide more results of our calculations for other symmetry directions in the 
Brillouin zone and some additional comparisons with ARPES experiments at different polarizations.

In some sense MXM direction shown in Fig. 1 is the simplest one among others since only few bands are present here.
The most intensive region here is around X point. For $k_z=\pi$ $3z^2-r^2$ band goes a bit down in energy (panel f). It
leads to lowering of intensity around X point for theoretical spectral function (panel e) and quantitatively reproduce
ARPES data at 160 eV (panel d).

In Fig. 2 for X$\Gamma$X high symmetry direction qualitative picture of bands evolution from $k_z=0$ to $k_z=\pi$
is the same as for M$\Gamma$M direction (see Fig. 5 in the main text).
Again the ARPES data at 80 eV (panel c) agrees better with $k_z=0$ LDA+DMFT results (panels a,b). Most intensive spots
of spectral function are formed at the crossing of $xz,yz$ branches at -0.1 eV. While for $k_z=\pi$ most intensive 
region appears around -0.2 eV, where $3z^2-r^2$ and $xz,yz$ bands are dominating.

To discuss different Fe-3d orbitals contribution to spectral function maps we used
experimental ARPES spectral functions obtained for different polarizations [1].
In Fig. 3 panel a corresponds to vertical polarization ARPES data in the M$\Gamma$M high symmetry direction
taken at 160 eV and panel f -- to horizontal polarization. For vertically polarized beam ``cap''--like structure
around $\Gamma$-point is formed mainly by $xz,yz$ orbitals (panels b and d for LDA+DMFT results). Surprisingly
the intensity of $xy$ band (panels c and e) is quite low in ARPES data and even not addressed in Ref. [1].
The $k_z$ dispersion of these bands near the Fermi level is almost absent.

Horizontally polarized beam (panel f) wipes out  $3z^2-r^2$ band forming ``M''--like structure around -0.2 eV.
For $k_z=\pi$ it has higher intensity than for $k_z=0$. It is in better agreement with to 159 eV data.

\begin{figure*}
\includegraphics[clip=true,width=\textwidth]{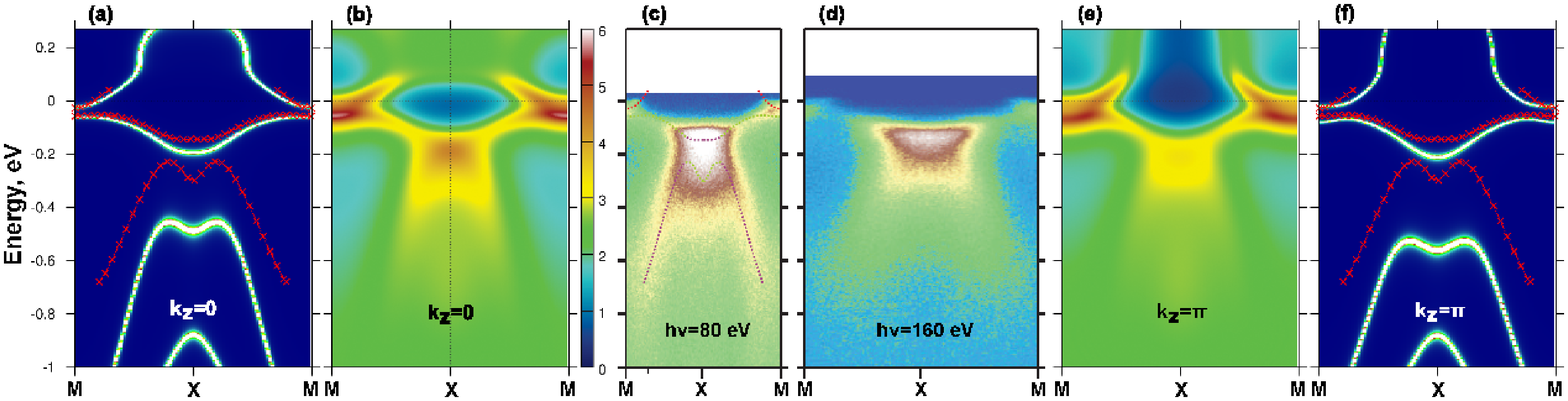}
\caption{Fig. 1. Comparison of experimental ARPES (panels c and d) [1] and LDA+DMFT (panels b and e)
spectral functions in the MXM high symmetry direction for NaFeAs near the Fermi level.
On the panels a and f experimental (crosses) [1] and theoretical (white lines) maxima dispersions 
of spectral functions are presented. The Fermi level $E_F$ is at zero energy.} 
\end{figure*}

\begin{figure*}
\includegraphics[clip=true,width=\textwidth]{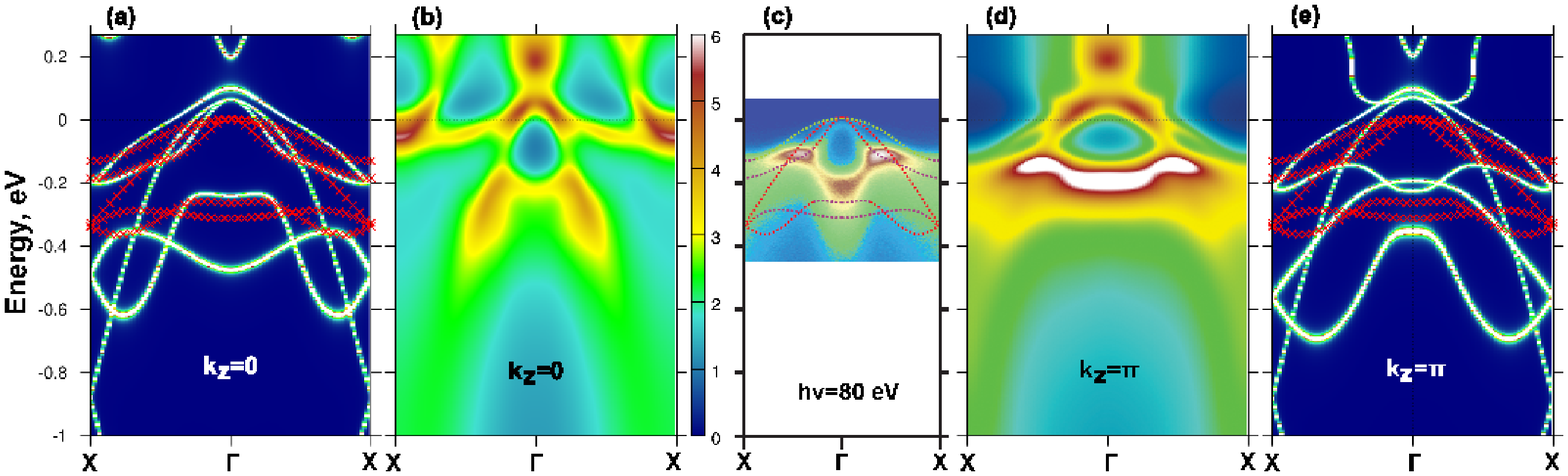}

\caption{Fig. 2. Comparison of experimental ARPES (panels c and d) [1] and LDA+DMFT (panels b and e)
spectral functions in the X$\Gamma$X high symmetry direction for NaFeAs near the Fermi level.
On the panels a and e experimental (crosses) [1] and theoretical (white lines) maxima dispersions 
of spectral functions are presented. The Fermi level $E_F$ is at zero energy.} 
\end{figure*}

\begin{figure*}

\includegraphics[clip=true,width=\textwidth]{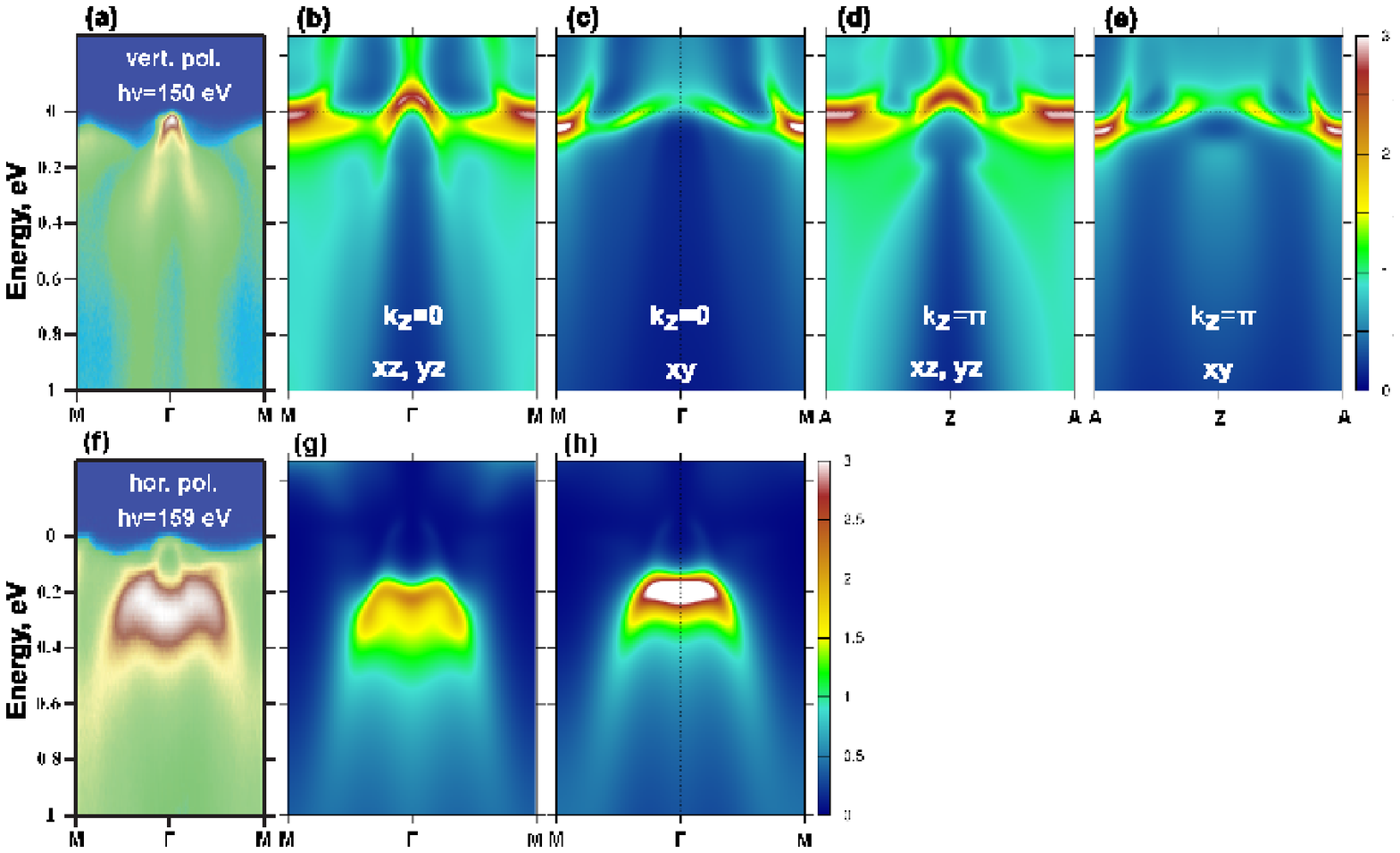}

\caption{Fig. 3. Comparison of experimental ARPES spectral functions with different polarization
(panel a -- vertical polarization, panel f -- horizontal polarization) [1]
and LDA+DMFT spectral functions for different Fe-3d orbitals: panels b-e -- $xz$,$yz$ and $xy$ contributions,
panels g,h -- $3z^2-r^2$ contribution in the M$\Gamma$M high symmetry direction for NaFeAs near the Fermi level
for $k_z=0$ and $k_z=\pi$ cases.
The Fermi level $E_F$ is at zero energy. } 
\end{figure*}

%

\end{document}